%% file: arxiv.tex
\documentclass[final]{article}

\usepackage{arxiv}

\usepackage[utf8]{inputenc} 
\usepackage[T1]{fontenc}    
\usepackage{hyperref}       
\usepackage{url}            
\usepackage{booktabs}       
\usepackage{amsfonts}       
\usepackage{nicefrac}       
\usepackage{microtype}      
\usepackage{lipsum}
\usepackage{graphicx}

\usepackage{xcolor}
\usepackage{tikz}
\usepackage{amsmath}
\usepackage{amssymb}

\usepackage{lineno}
\usepackage{algorithm2e}
\usepackage{textcomp} 
\usepackage{subcaption}
\usepackage{adjustbox}
\usepackage{booktabs}

\modulolinenumbers[5]
\usetikzlibrary{calc}
\usetikzlibrary{intersections}
\usetikzlibrary{through}

\title{Wind Flow Estimation in Thermal Sky Images for Sun Occlusion Prediction}

\author{
 Guillermo Terr\'en-Serrano \\
  Department of Electrical and Computer Engineering \\
  The University of New Mexico \\
  Albuquerque, NM 87131, United States\\
  \texttt{guillermoterren@unm.edu} \\
 \And
  Manel Mart\'inez-Ram\'on \\
  Department of Electrical and Computer Engineering \\
  The University of New Mexico \\
  Albuquerque, NM 87131, United States\\
  \texttt{manel@unm.edu} \\
}

\begin{document}

\maketitle

\begin{abstract}
    Moving clouds affect the global solar irradiance that reaches the surface of the Earth. As a consequence, the amount of resources available to meet the energy demand in a smart grid powered using Photovoltaic (PV) systems depends on the shadows projected by passing clouds. This research introduces an algorithm for tracking clouds to predict Sun occlusion. Using thermal images of clouds, the algorithm is capable of estimating multiple wind velocity fields with different altitudes, velocity magnitudes and directions. 
\end{abstract}


\section{Introduction}

The portion of energy generated by PV systems is increasing in exponential scale since 2000 \cite{PVgrowth}. To continue this trend, it is important to provide a reliable energy supply \cite{BEYEA2010}. An algorithm that forecasts Sun occlusion equips a grid with the capability of efficiently controlling the dispatch and storage of energy \cite{CHEN2020}.

The forecasting horizon required to nowcast Sun occlusion is between 1 to 5 minutes ahead \cite{VOYANT2017}. Numerical weather prediction models that use mesoscale meteorology have problems of collinearity \cite{GARCIA2018} and the forecast is not effective within the required range for nowcasting. Satellite imaging systems are practical when the horizon ranges from 15 minutes to an hour \cite{ARBIZU2017}. An alternative to satellite imaging systems are ground-based all-sky imagers \cite{TSI2001}. Visible light cameras are inexpensive, and a lens or concave mirror may be attached to increase the Field Of View (FOV) \cite{CHENG2017}. The disadvantage of visible light imaging is that the pixels in the circumsolar region are saturated. This is especially problematic for nowcasting Sun occlusions. Ground-based thermal sky-imaging systems reduce the saturation of the pixels in the circumsolar area \cite{MAMMOLI2019}. These systems have been used to measure spatiotemporal cloud statistics to establish an optical link in Earth-space communications \cite{NUGENT2009}.

Previous investigations have set the precedent of using ground-based sky-images in computer vision algorithms to detect clouds and to estimate their motion \cite{CHOW2011}. The Kalman filter has been used to track clouds and approximate their pathlines \cite{CHENG2017}. Artificial neural networks \cite{KONG2020} and Support Vector Machines (SVM) \cite{DENG2019} have been shown capable of finding spatiotemporal correlations between solar irradiance and sky-images of clouds.

This investigation introduces an algorithm which predicts the pathlines of clouds moving in different wind velocity fields. A Multi-Output Weighted Support Vector Machine with Flow Constraints ($\varepsilon$-MO-WSVM-FC) is proposed to estimate the multiple wind velocity fields detected in an image. The velocity vectors are computed using a weighted implementation of the Lucas-Kanade (WLK) algorithm. The weights are the posterior probabilities of the temperatures in an image inferred using a Beta Mixture Model (BeMM). The velocity vectors are segmented and subsampled so that the implementation of the proposed cloud tracking algorithm is feasible for nowcasting Sun occlusions.

\section{Dataset}

The data used in this investigation was acquired with a sky imager utilizing a solar tracker which maintains the Sun in the center of the images. The sky imager is equipped with a Lepton 2.5 thermal camera that measures temperature in centi-kelvin degrees. The resolution of the camera is $80 \times 60$ pixels and the diagonal FOV is $60^\circ$. The thermal sky-imaging system is located at the UNM-ECE building. The weather parameters are measured by a weather station at the UNM Hospital.

\subsection{Thermal Sky Images}

The intensity of a pixel $i,j$ in a thermal image is a temperature measurement in centi-kelvin degrees \cite{TERREN2020a}. The temperatures of the pixels in an image are defined as $\mathbf{T}^k = \{ T^k_{ij} \in \mathbb{R} \mid \forall i = 1, \ldots, M, \ \forall j = 1, \ldots, N \}$. The heights of the pixels are defined as $\mathbf{H}^k = \{ H_{ij} \in \mathbb{R} \mid \forall i = 1, \ldots, M, \ \forall j = 1, \ldots, N \}$. The heights of the pixels are calculated using a linear function defined as $\phi : (T^k_{ij}, \Gamma_{MARL}, T^{air}) \mapsto H^k_{ij}$ which depends on the temperature of the pixel $T_{ij}$, the Moist Adiabatic Lapse Rate (MARL) \cite{MURALIKRISHNA2017} and the air temperature at the ground-level $T^{air}$.

\subsection{Cloud Velocity Vectors}

The distribution of the temperatures in a sky-image with multiple layers of clouds is inferred using a BeMM. The number of cloud layers in an image is defined as $C^k$, and is determined by a previously trained algorithm. When clouds are flowing in different wind velocity fields, the BeMM is expected to have $C^k$ clusters.

\subsubsection{Beta Mixture Model}

Consider the temperatures $\bar{T}_{ij}$ of a given image (omitting superindex $k$). The distribution of the normalized temperatures can be approximated by a mixture of beta distributions $\bar{T} \sim Be ( \alpha_c, \beta_c )$ with the density function,
\begin{align}
    \label{eq:beta_distribution}
    f \left( \bar{T}_{ij} ; \alpha_c, \beta_c \right) = \frac{1}{\mathrm{B} \left( \alpha_c, \beta_c \right)} \cdot \bar{T}_{ij}^{\alpha_c - 1} \cdot \left( 1 - \bar{T}_{ij}\right)^{\beta_c - 1}
\end{align}
where $\bar{T}_{ij} \in (0, 1)$,  the beta function is $\mathrm{B} ( \alpha_c, \beta_c ) = [ \Gamma ( \alpha_c ) \Gamma ( \beta_c ) ] / [\Gamma ( \alpha_c + \beta_c ) ] $, and the gamma function is $\Gamma ( \alpha_c ) = ( \alpha_c - 1 )!$.


The parameters in the clustering of beta distributions can be directly computed applying the Expectation Maximization (EM) algorithm \cite{MURPHY2012}. In the E stage, a posterior $\gamma_{ijc} \triangleq p(y_{ij} = c \mid \bar{T}_{ij}, \boldsymbol{\theta})$ can be assigned to each sample from prior probabilities for the classes an the likelihoods \eqref{eq:beta_distribution}. In the M stage, the parameters $\alpha_c$ and $\beta_c$ of each cluster that maximize the log-likelihood are computed by gradient descent.


The cloud average heights in a frame are computed as 
\begin{align}
    \hat{H}_c = \frac{\sum_{ij} \gamma_{ijc} \cdot H_{ij} \cdot \mathbb{I} \left( b_{ij} = 1 \right)}{\sum_{ij} \gamma_{ijc} \cdot \mathbb{I} \left( b_{ij} = 1 \right)},
\end{align}
where $\mathbb{I} \left( \cdot \right)$ is the indicator function. An image segmentation algorithm indicates which pixels belong to a cloud, so that $\mathbf{B} = \{ b_{ij} \in \mathbb{B} \mid \forall i = 1, \ldots, M, \ \forall j = 1, \ldots, N \}$ is a binary image where 0 is a clear sky pixel, and 1 is a pixel belonging to a cloud \cite{TERREN2020c}. 


\subsubsection{Weighted Lucas-Kanade}

The method implemented to compute the velocity vectors is WLK \cite{SIMON2003}. The weights $\gamma_{ijc}$ are the posterior probabilities of the BeMM. A pixel $i,j$ has a velocity vector for each cloud layer $c$ in a frame. The optimal window size, weighted least-squares regularization, and differential kernel amplitude are: $\mathcal{W} = 16 \ [ \mathrm{pixels}^2 ]$, $\tau = 1 \times 10^{-8}$, and $\sigma = 1$ respectively. The velocity components are $\text{u}_{ijc} $ and $\text{v}_{ijc}$. The velocity vectors in pixels per frame  are transformed to m/s using the geospatial transformation of the perspective  \cite{TERREN2020a}, which is a function of the Sun's elevation and azimuth angles $\psi : \left( \varepsilon, \alpha \right) \mapsto \Delta \mathbf{x}_{ij} $ 
The transformations are
\begin{equation}
    \label{eq:geometric_transformation}
    \begin{split}
        u_{ij} &= \frac{\delta}{f_r}  \cdot \Delta x_{ij} \sum_{c = 1}^C  \hat{H}_c \cdot \gamma_{ijc} \cdot \text{u}_{ijc}  \\
        v_{ij} &= \frac{\delta}{f_r} \cdot \Delta y_{ij} \sum_{c = 1}^C \hat{H}_c \cdot \gamma_{ijc} \cdot \text{v}_{ijc}
    \end{split}
\end{equation}
where $f_r$ is the frame rate and $\delta$ is the scale of the velocity vectors.

\subsubsection{Velocity Vector Segmentation}

The pixel intensity difference between two consecutive frames is computed to find which percentage of pixels $\tau$ show more change. The wind velocity field is approximated using the segmented velocity vectors of $\ell$ last frames. Hence, the set of velocity vectors available to compute the wind velocity field are,
\begin{align}
    \label{eq:dataset}
    \tilde{\mathbf{V}}^{k} = \left[ {\begin{array}{ccc}
       {{\bf V}'}^k \\
       \vdots \\
       {{\bf V}'}^{k - \ell} \\
      \end{array} } \right] \in \mathbb{R}^{2 \times N^k},
\end{align}
the number of samples in $\tilde{\mathbf{V}}^k$ is $N^k$, this number is not the same in each frame $k$. 

\subsubsection{Inference of Velocity Vector and Height Distributions}

A velocity vector $\tilde{\mathbf{v}}_i$ (omitting superindex $k$) in the set $\tilde{\mathbf{V}}^k = \{ \tilde{\mathbf{v}}^k_i \in \mathbb{R}^2 \mid \forall i = 1, \dots, N^k\}$ is assumed to belong to a cloud layer $c$. The probability of a vector to belong to a cloud layer $c$ is modelled as an independent normal random variable $\tilde{\mathbf{v}}_i \sim p \left( \tilde{\mathbf{v}}_i \mid  \boldsymbol{\mu}_c, \boldsymbol{\Sigma}_c \right)=\mathcal{N}( \boldsymbol{\mu}_c, \boldsymbol{\Sigma}_c )$. 
In the case when two cloud layers were detected, we propose to infer the probability distribution of velocity vectors' in each cloud layer as
\begin{align}
    p \left( \tilde{\mathbf{v}}_i \mid \boldsymbol{\Theta} \right) \propto p \left( \tilde{\mathbf{v}}_i \mid \boldsymbol{\mu}_1, \boldsymbol{\Sigma}_1 \right)^{\lambda_i} \cdot  p \left( \tilde{\mathbf{v}}_i  \mid \boldsymbol{\mu}_2, \boldsymbol{\Sigma}_2 \right)^{\left(1 - \lambda_i \right)},
\end{align}
where $\boldsymbol{\Theta} = \{ \boldsymbol{\lambda}, \boldsymbol{\mu}_1, \boldsymbol{\Sigma}_1, \boldsymbol{\mu}_2, \boldsymbol{\Sigma}_2\}$, and $\lambda_i \in \{0, 1\}$. The log likelihood is then expressed as the linear combination,
\begin{align}
	\label{eq:velocity_vectors_distribution}
    \log p ( \tilde{\mathbf{v}}_i \mid {\boldsymbol{\Theta}} ) \propto \lambda_{i1} \log p ( \tilde{\mathbf{v}}_i \mid {\boldsymbol{\mu}}_1, {\boldsymbol{\Sigma}}_1 ) + \lambda_{i2} \log p ( \tilde{\mathbf{v}}_i  \mid {\boldsymbol{\mu}}_2, {\boldsymbol{\Sigma}}_2 ),
\end{align}
where $\lambda_{i1} = \lambda_i$ and $\lambda_{i2} = 1 - \lambda_i$. The probabilistic model parameters are inferred using a fixed-point variation of the Iterated Conditional Modes (ICM) \cite{BESAG1986}. After completing the inference of the velocity vectors distribution, it is possible to infer the cloud layer's height using the same method and a likelihood $H_{i,j} \sim \mathcal{N} (\mu_c, \sigma^2_c) $.

\subsubsection{Velocity Vector Subsampling}

To reduce the computational burden of the algorithm, a subset of $N^*$ velocity vectors is selected according to the estimated probability distributions of the vectors $p ( \tilde{\mathbf{v}}^{*k}_i \mid \boldsymbol{\theta}_c )$, their posterior probabilities are $z_{i}^{*k}$.

\section{Wind Velocity Field Approximation}

Three methods were implemented to estimate the extrapolation function and compare their performances. The first method uses a weighted $\varepsilon$-support vector regression machine  ($\varepsilon$-WSVM) for each one of the velocity components. The second method is a $\varepsilon$-MO-WSVM that estimates both velocity components. The third is an innovation which uses a $\varepsilon$-MO-WSVM with flow constraints ($\varepsilon$-MO-WSVM-FC) to estimate both velocity components. The flow constraints are used to force the extrapolated wind flow to have zero divergence or curl, so it can be assumed that, in the approximated wind flow, streamlines are equivalent to the cloud pathlines.

\subsection{$\varepsilon$-WSVM}

The regression problem can be formulated as the optimization of a function with the form,
\begin{equation}
    f(\mathbf{x}_i) = \mathbf{w}^\top \varphi ( \mathbf{x}_i ) + b, \ \forall i = 1, \dots, N^*, \ \mathbf{w}, \mathbf{x}_i \in \mathbb{R}^D, \ b \in \mathbb{R},
\end{equation}
where $\mathbf{x}_i \triangleq \mathbf{x}^{*k}_i$ and $\varphi(\cdot)$ is a transformation into a higher dimensional (possibly infinite) Hilbert space $\mathcal{H}$ endowed with a dot product $\mathcal{K}({\bf x}_i,{\bf x}_j)=\langle \varphi({\bf x}_i),\varphi({\bf x}_j)\rangle$. 

\subsubsection{Support Vector Machine for Regression}

Assuming $\mathbf{v}_i =\{u_i,v_i\} \triangleq \mathbf{v}_i^{*k}$, the regression problem in a $\varepsilon$-SVM is formulated
\cite{SCHOLKOPF2000}, where the samples are weighted by their probability of belonging to wind velocity field $c$,
\begin{align}
    z_i \triangleq z_i^{*k}, \quad 
    c_i = z_i \cdot \frac{C}{N}, \quad z_i \in \mathbb{R}^{\leq 1},
\end{align}
leading to a weighted SVR whose primal formulation is
\begin{align}
    \min_{\mathbf{w}, b, \xi, \xi^*} & \quad \frac{1}{2} \| \mathbf{w} \|^2 + \frac{C}{N} \sum_{i = 1}^N z_i \left( \xi_i + \xi_i^* \right) \label{eq:primal}\\ 
    \mathrm{s.t.} & 
    \begin{cases}
    u_{i}-\mathbf{w}^\top \varphi \left( \mathbf{x}_i \right) - b & \leq \varepsilon + \xi_i \\
    \mathbf{w}^\top \varphi \left( \mathbf{x}_i \right) + b  - u_{i} & \leq \varepsilon + \xi_i^* \\
   \xi_i, \xi_i^* & \geq 0
    \end{cases}  \ i = 1, \ldots, N, \label{eq:primal_constraints}
\end{align}
and identically for $v_i$. 
We used linear, square exponential and polynomial kernels \cite{SHAWE2004} in the experiments. The dual problem formulation for the SVR and its solution can be found in \cite{SMOLA2004}.

\subsubsection{Multi-Output Weighted Support Vector Machine}

When the wind velocity field function is approximated by $\varepsilon$-MO-SVM,the primal regression can be formulated as 
\begin{equation}
    {\bf v}_i = {\bf W}^\top\varphi({\bf x}_i)+{\bf b},
\end{equation}
where each one of the column vectors of primal parameter matrix ${\bf W}$ approximates one of the velocities in vector ${\bf y}_i$. Primal parameters are a function of the dual parameters as well, but the dual parameters $\boldsymbol{\alpha}_i, \boldsymbol{\alpha}_i^{*}$ are vectors in a $2$-dimensional multi-output problem. 

Since independent variables are represented in vectors ${\bf v}_i$, the training set is defined in a vector  $\tilde{\mathbf{v}}_{1 \times 2N}$, and so are the dual parameters $\tilde{\boldsymbol{\alpha}}_{1 \times 2N}$ and $\tilde{\boldsymbol{\alpha}}^*_{1 \times 2N} $ for notation simplicity. The optimization for this model is similar to the one for the standard SVR.

\subsubsection{Multi-Output Weighted Support Vector Machine with Flow Constraints}

Assuming that the analyzed air parcel is sufficiently small so that the flow can be considered approximately incompressible and irrotational, a new set of flow constraints are added to the original set of constraints with the purpose of visualizing the wind velocity field to force the divergence and the vorticity to zero:
\begin{align}
    \label{eq:flow_constraints}
    \mathrm{s.t.}\begin{cases} 
        \left( \tilde{\mathbf{v}}^{k\top}_c \mathbf{\Delta}_{xy} \mathbf{\dot{V}} \right)  \cdot \left( \tilde{\mathbf{v}}^{k\top}_c \mathbf{\Delta}_{xy} \mathbf{\dot{V}} \right)^\top &= 0 \\
        \left(  \tilde{\mathbf{v}}^{k\top}_c \mathbf{\Delta}_{xy} \mathbf{\dot{D}} \right) \cdot \left( \tilde{\mathbf{v}}^{k\top}_c \mathbf{\Delta}_{xy} \mathbf{\dot{D}} \right)^\top &= 0.
    \end{cases} 
\end{align}
To compute the vorticity and divergence, the differentiation of the velocity field along the x-axis, the and y-axis is implemented using operator
\begin{align}
    \label{eq:matrix_1}
    \mathbf{\Delta}_{xy} =
    \begin{bmatrix}
        \boldsymbol{\Delta}_x & \mathbf{0} \\
        \mathbf{0} & \boldsymbol{\Delta}_y
    \end{bmatrix}_{2N \times 2N}.
\end{align}
The differential operators are $\boldsymbol{\Delta}_x = \mathbf{L} - \mathbf{I}$ and $\boldsymbol{\Delta}_y = \mathbf{L}^\prime - \mathbf{I}$ $\forall ij \in \{1,\dots, N\}$, where $\mathbf{L}_{i,j} = \delta_{ij + 1}$ and $\mathbf{L}^\prime_{i,j} = \delta_{ij + M + 1}$ are a lower sift matrices, $\delta_{ij}$ is a Kronecker delta function and $\mathbf{I}_{N \times N}$ is a diagonal matrix. The operators of the velocity field's vorticity and divergence are $\mathbf{\dot{V}} = \begin{bmatrix} \mathbf{I} & \mathbf{I} \end{bmatrix}^\top$ and
$\mathbf{\dot{D}} = \begin{bmatrix} \mathbf{I} & -\mathbf{I} \end{bmatrix}^\top$ respectively.


The velocity field is extrapolated to the entire frame using the inferred parameters in frame $k$
\begin{align}
    \label{eq:extrapolation}
    \hat{\mathbf{v}}^k_c = \left( \tilde{\boldsymbol\alpha}^k_c - \tilde{\boldsymbol\alpha}^{*k}_c \right) \cdot \mathcal{K} \left( \mathbf{X}^{*k}, \mathbf{X} \right) + \mathbf{b}_c^k,
\end{align}
where the velocity components are $\hat{\mathbf{U}}^k_c \triangleq \hat{\mathbf{v}}^k_{xc}$, and $\hat{\mathbf{V}}^k_c \triangleq \hat{\mathbf{v}}^k_{yc}$, where  $\hat{\mathbf{U}}_c^k, \hat{\mathbf{V}}^k_c \in \mathbb{R}^{M \times N} $. 
To compute the flow constraints, the velocity field has to be extrapolated to the whole frame using Eq. \eqref{eq:extrapolation}. The constraints in Eq. \eqref{eq:flow_constraints} are added to the constraints in the dual formulation of the $\varepsilon$-MO-WSVM.

\subsection{Streamlines}


A wind velocity field may have divergence and vorticity. However, the air parcel in one frame is very small compared to the whole volume of air contained in the atmosphere. Within this frame we assume that there is no divergence or vorticity in the approximated wind velocity field, and that  streamlines are equivalent to the pathlines.

The trapezoidal rule of numerical analysis is applied to solve the definite integrals \cite{DRAGOMIR1999}. The values of a streamline $\boldsymbol{\Phi}_c^k$ and a potential line $\boldsymbol{\Psi}_c^k$ are (omitting superindex $k$),
\begin{equation}
    \begin{split}
    \boldsymbol{\Phi}_c &= \frac{\hat{H}_c}{2} \left[ \left\{ \sum^i_{m} \hat{\bf u}_{mc} \odot \Delta {\bf y}_{mc} \right\}^M_{i} - \left\{ \sum^j_{m} \hat{\bf v}_{mc} \odot \Delta {\bf x}_{mc} \right\}^N_{j} \right] \\
    \boldsymbol{\Psi}_c &= \frac{\hat{H}_c}{2} \left[ \left\{ \sum^j_{m} \hat{\bf u}_{mc} \odot \Delta {\bf x}_{mc} \right\}^N_{j} + \left\{ \sum^i_{m} \hat{\bf v}_{mc} \odot \Delta {\bf y}_{mc}  \right\}^M_{i}  \right]
    \end{split}
    \label{eq:streamlines}
\end{equation}
where $\odot$ denotes the element-wise matrix multiplication.


\section{Results and Discussion}

The algorithm validation has two steps. The first step is the parameter validation of the algorithm which computes and selects the velocity vectors. In the second step, the parameters of the $\varepsilon$-MO-WSVM and $\varepsilon$-MO-WSVM-FC are validated. The dataset is divided into training and testing. The training set is used in both validation steps. The training set is composed of sequences of 21 consecutive frames from 6 different days. The selection of image sequences was based on the variety of different types of clouds distributed across different heights. In particular, 3 of the sequences contain a single layer of clouds, and the other 3 contain multiple layers of clouds.

\begin{figure}[ht]
    \begin{subfigure}{\linewidth}
        \centering
        \includegraphics[scale = 0.4]{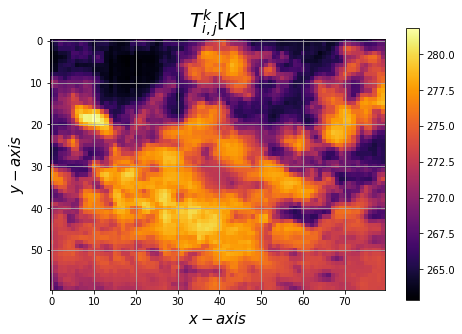}
        \includegraphics[scale = 0.4]{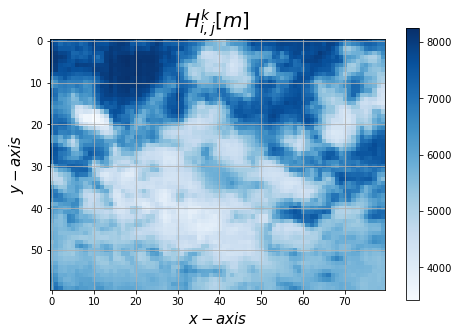}
    \end{subfigure}
    \begin{subfigure}{\linewidth}
        \centering
        \includegraphics[scale = 0.4]{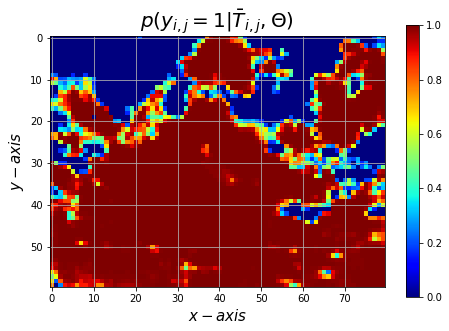}
        \includegraphics[scale = 0.4]{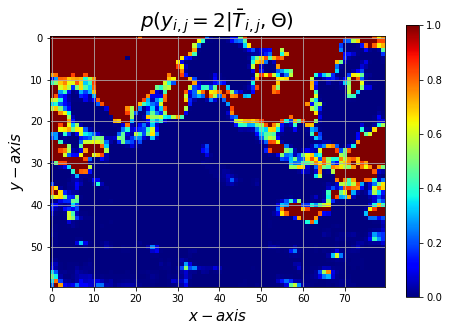}
    \end{subfigure}
    \caption{The images in the first row shows the temperature (left) and the height (right) of the pixels. The images in the second row shows the posterior probabilities of the BeMM.}
    \label{fig:wind_velocity_field}
\end{figure}

The parameters validated in the algorithm for the computation and selection of velocity vectors are $\delta$, $\tau$, $\ell$ and $N^*$. The parameter validation of the algorithm requires labeling the approximated wind velocity field. The necessary labels to define the wind velocity field in a frame are: height, velocity magnitude and angle. The clouds in a layer were segmented to compute their average height, and the pathline which intercepts the Sun was manually segmented to calculate the distance that a cloud is moving as well as its direction.

\begin{figure}[htb]
    \begin{subfigure}{0.475\linewidth}
        \centering
        \includegraphics[scale = 0.4]{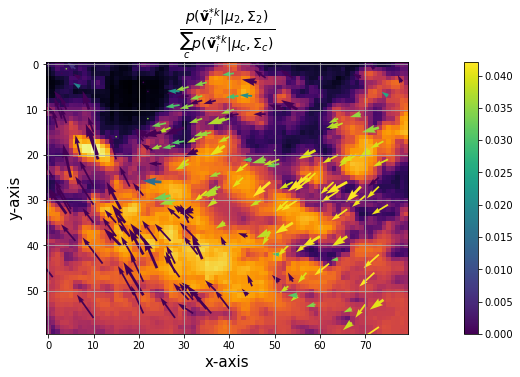}
    \end{subfigure}
    \begin{subfigure}{0.475\linewidth}
        \centering
        \includegraphics[scale = 0.4]{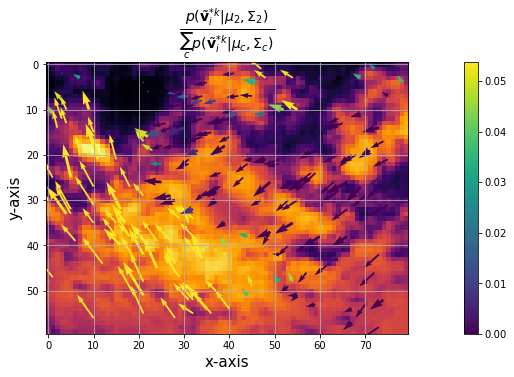}
    \end{subfigure}
    \caption{The images show the selected velocity vectors in their respective coordinates. The color intensity represent the velocity vectors posterior probability of belong to the upper layer (left) or to the lower layer (right).}
\label{fig:velocity_samples}
\end{figure}

The wind velocity field was approximated in each frame of the training set for each set of parameters validated $\Theta = \{ \delta, \tau, \ell, N^* \}$. The approximation was performed using an independent linear $\varepsilon$-WSVMs for each velocity component. The parameters $C$ of the $\varepsilon$-WSVMs were cross-validated in each training frame. The dataset used in the parameter validation of the $\varepsilon$-SVMs does not require labels. The targets are the velocity vectors computed using the WLK. The algorithm is trained online in each new frame. The $N^*$ selected velocity vectors are divided in training 75\% and testing 25\% sets. This training set is used to validate the parameters of the $\varepsilon$-WSVMs and to train the model. The testing set is used to evaluate the Weighed Mean Absolute Error (WMAE) achieved by the model.

\input{table_arxiv}

The Mean Absolute Percentage Error (MAPE) was computed using the labels and the approximated average height, magnitude and angle of the wind velocity field. A MAPE was calculated in each frame of the training set. The aim was to find the set of parameters with less MAPE but without considerable variation between consecutive frames. To meet this end, two metrics were calculated and averaged together. The first metric is computed by averaging the MAPE obtained in each training frame. The second metric is computed averaging together the differences of MAPE between consecutive frames in the training set. The optimal validation parameters of the algorithm are $\delta = 2.29$, $\tau = 0.95$, $\ell = 6$ and $N^* = 200$. The implementation of the algorithm for the computation and selection of velocity vectors is shown in Fig \ref{fig:wind_velocity_field}-\ref{fig:velocity_samples}. The extracted features and the weights used to compute the velocity vectors with the WLK method are shown in Fig. \ref{fig:wind_velocity_field}. The selected velocity vectors used to approximate a multilayer wind flow are shown in Fig. \ref{fig:velocity_samples}.

\begin{figure}[htb]
    \begin{subfigure}{\linewidth}
        \centering
        \includegraphics[scale = 0.25]{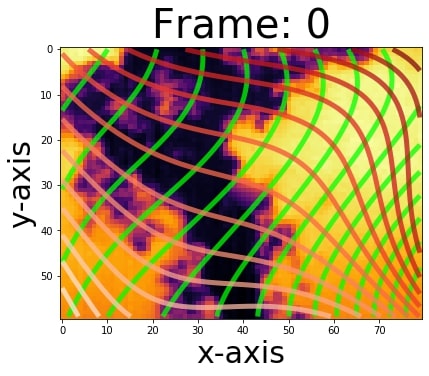}
        \includegraphics[scale = 0.25]{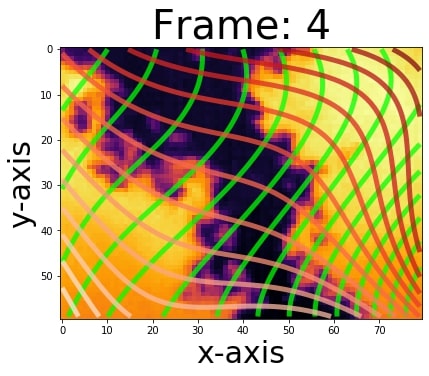}
        \includegraphics[scale = 0.25]{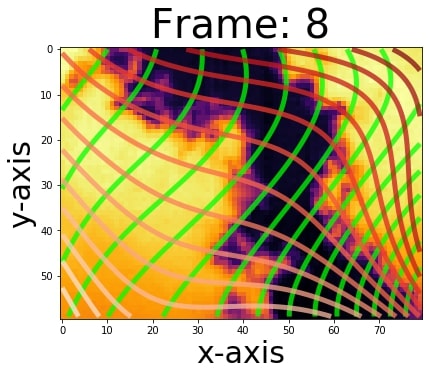}
        \caption{Wind flow approximated with $\varepsilon$-MO-WSVM using a $\mathcal{P}^3$ kernel.}
        \label{fig:day_2_layer_0_turbulent}
    \end{subfigure}
    \begin{subfigure}{\linewidth}
        \centering
        \includegraphics[scale = 0.25]{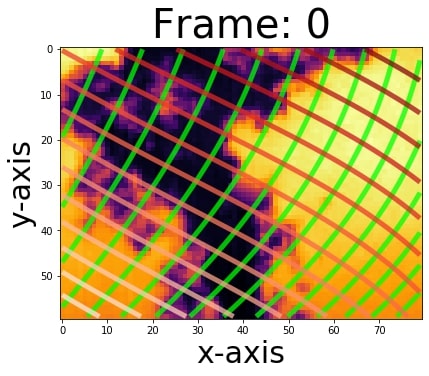}
        \includegraphics[scale = 0.25]{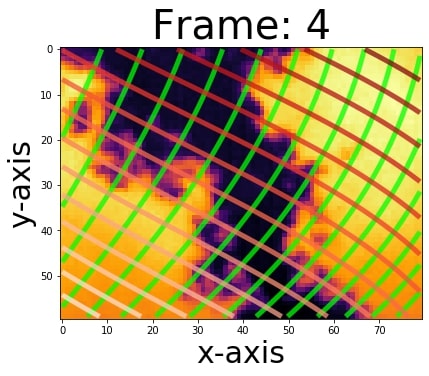}
        \includegraphics[scale = 0.25]{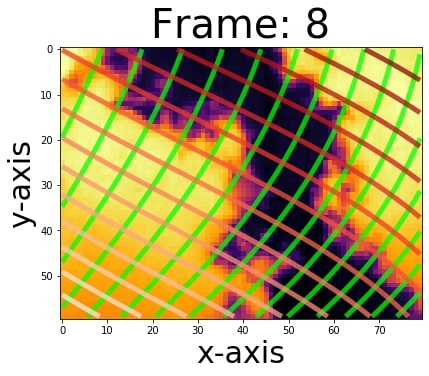}
        \caption{Wind flow approximated with $\varepsilon$-MO-WSVM-FC using a linear kernel.}
        \label{fig:day_2_layer_0}
    \end{subfigure}
    \caption{The thermal images are organized in time from left to right. The first images is when the streamlines (green) and potential lines (red) were calculated. The second and third images are after 1 minutes and 2 minutes respectively. }
\end{figure}

The testing set is composed of 10 sequences of 21 consecutive images. 5 of the sequences have one layer of clouds, and the other 5 have multiple layers of clouds. The distribution of the clouds is different in each one of the sequences and each sequence was recorded during a different hour and day. The performances of the $\varepsilon$-WSVMs were evaluated using the testing subset. Two experiments were performed with each of the $\varepsilon$-WSVMs. In the first experiment, the parameters of the $\varepsilon$-WSVMs are validated in each testing frame. In the second experiment, the parameters are fixed in each testing frame to the optimal parameters computed with the training set. Using the optimal parameters for the computation and selection of velocity vectors algorithm, the parameters of the $\varepsilon$-WSVMs were validated in each training frame. The optimal parameters of the $\varepsilon$-WSVMs are the result of averaging together the validated parameters in each training frame.

\begin{figure}[htb!]
    \begin{subfigure}{\linewidth}
        \centering
        \includegraphics[scale = 0.25]{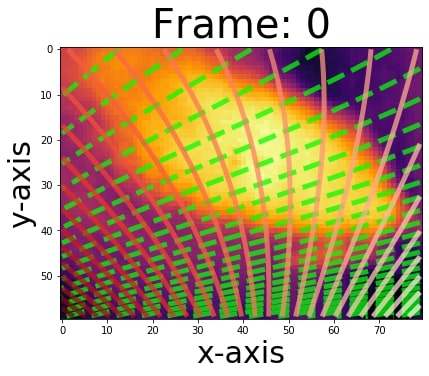}
        \includegraphics[scale = 0.25]{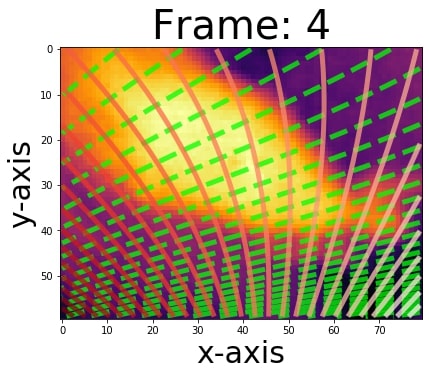}
        \includegraphics[scale =     0.25]{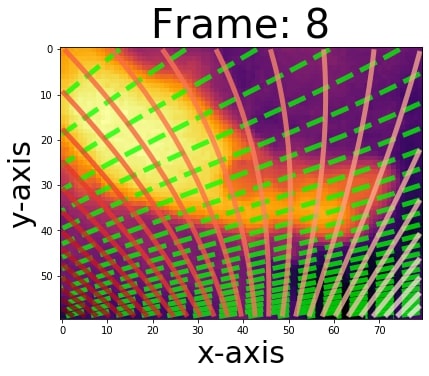}
        \caption{Upper layer of clouds Streamlines (green) and potential lines (red).}
        \label{fig:day_5_layer_0}
    \end{subfigure}
    \begin{subfigure}{\linewidth}
        \centering
        \includegraphics[scale = 0.25]{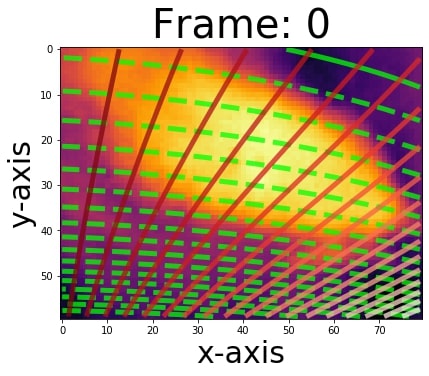}
        \includegraphics[scale = 0.25]{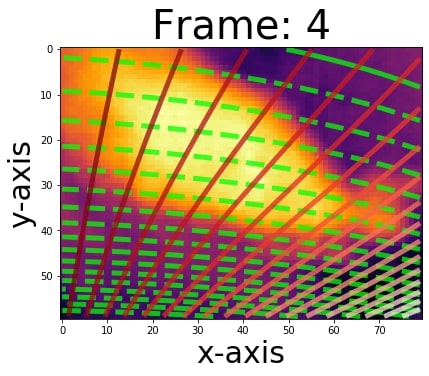}
        \includegraphics[scale = 0.25]{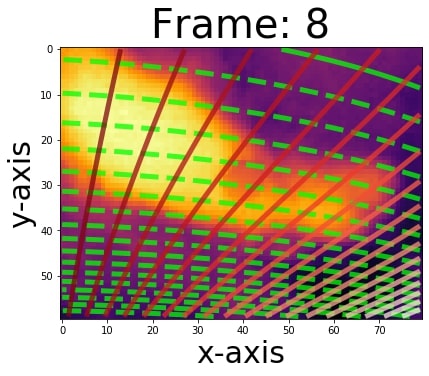}
        \caption{Lower layer of cloud Streamlines (green) and potential lines (red).}
        \label{fig:day_5_layer_1}
    \end{subfigure}
    \label{fig:multilayer_flow}
    \caption{The thermal images are organized sequentially from left to right. When the wind flow was approximated is frame 0, after 1 minute is frame 4 and after 2 minutes is frame 8.}
\end{figure}

When the samples are weighted, the performance of the models increases (see Table \ref{tab:wsvm_fc}). Wind flows approximated using a $\varepsilon$-MO-WSVM have low WMAE and computing time but high divergence and vorticity (see Table \ref{tab:wsvm_fc}). The wind flows approximated using $\mathcal{P}^2$ or $\mathcal{P}^3$ kernels are very turbulent (see Fig. \ref{fig:day_5_layer_0}). When the wind flow is approximated using the $\varepsilon$-MO-WSVM-FC with linear and RBF kernels, the wind flow has low divergence and vorticity (see Fig. \ref{fig:day_5_layer_1}). In addition, the validation of the $\varepsilon$-MO-WSVM-FC requires less computing time when using the linear and RBF kernels (see Table \ref{tab:wsvm_fc}).

The most suitable method in the application of nowcasting requires a compromise between WMAE, computing time, divergence and vorticity. The computing time of a prediction has to be feasible for nowcasting, and negligible divergence and vorticity are required to use pathlines to approximate the streamlines. Taking this into consideration, the most suitable method is the $\varepsilon$-MO-WSVM-FC with a linear kernel. The approximated streamlines and potential lines using this method in a single layer flow are shown in Fig. \ref{fig:day_2_layer_0}, and in a multilayer wind flow in Fig. \ref{fig:day_5_layer_0}-\ref{fig:day_5_layer_1}.

The experiments were carried out in the Wheeler high performance computer of the UNM-CARC, which uses a SGI AltixXE Xeon X5550 at 2.67GHz with 6 GB of RAM memory per core, 8 cores per node.

\section{Conclusions}

The proposed algorithm uses features extracted from thermal images of clouds to estimate the wind velocity fields in which clouds are flowing. The velocity vectors are computed using the WLK method for each layer of clouds detected in the thermal images. The distribution of the velocity vectors and the cloud height
are used to infer which layer of clouds is lower and which is higher. The wind velocity field is extrapolated to the entire image using a $\varepsilon$-MO-WSVM-FC with a subset of the velocity vectors. The wind flow streamlines (i.e. pathlines) and potential lines are computed using the approximated wind velocity fields in an image.

Further research in this area will focus on the application of the cloud tracing algorithm in global solar irradiance forecasting, using the predicted pathlines. The pathline that will intercept the Sun may be used to anticipate when a cloud will obstruct the direct radiation of the Sun.

\section*{Acknowledgment}

Partially supported by NSF EPSCoR grant number OIA-1757207 and the King Felipe VI endowed Chair. Authors would like to thank the UNM Center for Advanced Research Computing (CARC) for providing the high performance computing and large-scale storage resources used in this work.

\bibliographystyle{unsrt}  
\bibliography{mybibfile}

\end{document}

%% file: table_arxiv.tex
\begin{table}
    \centering
    \small
    \setlength{\tabcolsep}{3.5pt}
    \renewcommand{\arraystretch}{1.5}
        \begin{tabular}{lccccccccccccc}
        \toprule
        {} & \multicolumn{4}{c}{Parameters} & \multicolumn{4}{c}{Cross-Validated} & \multicolumn{5}{c}{Fixed} \\
        $\mathcal{K} \left(\mathbf{x}, \mathbf{x}^*\right) $ & $C$ & $\varepsilon$ & $\gamma$ & $\beta$ & MAE & WMAE & $\nabla \cdot \vec{V}$ & $\nabla \times \vec{V}$ & MAE & WMAE & $\nabla \cdot \vec{V}$ & $\nabla \times \vec{V}$ & Time [s] \\
        \midrule
        {} &  \multicolumn{13}{c}{$\varepsilon$-MO-WSVM} \\
        \midrule
        Linear & 31.06 & 0.31 & & & 13.27 & \textbf{12.49} & \textbf{1.30}$\mathbf{\cdot 10^3}$ & \textbf{1.35}$\mathbf{\cdot 10^3}$ & 13.17 & \textbf{12.45} & \textbf{1052.43} & \textbf{1196.68} & 19.95 \\
        RBF & 38.71 & 0.36 & 17.81 & & 14.00 & 13.13 & 1.21$\cdot 10^3$ & 1.22$\cdot 10^3$ & 13.83 & 12.87 & 620.99 & 641.04 & 12.22 \\
        $\mathcal{P}^2$ & 34.73 & 0.28 & 4.65 & 10.66 & 14.25 & 13.53 & 1.43$\cdot 10^4$ & 1.71$\cdot 10^4$ & 17.69 & 16.89 & 5.45$\cdot 10^4$ & 6.54$\cdot 10^4$ & 32.49 \\
        $\mathcal{P}^3$ & 38.47 & 0.22 & 4.19 & 2.33 & 19.29 & 18.12 & 8.89$\cdot 10^5$ & 8.93$\cdot 10^5$ & 42.77 & 43.03 & 1.84$\cdot 10^6$ & 2.16$\cdot 10^6$ & 44.77 \\
        \midrule
        {} & \multicolumn{13}{c}{$\varepsilon$-MO-WSVM-FC} \\
        \midrule
        Linear & 38.50 & 0.19 & & & 13.47 & \textbf{13.31} & \textbf{0} & \textbf{0} & 13.57 & 13.32 & 0 & 0.01 & 55.53 \\
        RBF & 38.52 & 0.35 & 13.92 & & 14.09 & 13.72 & 24.48 & 24.25 & 13.65 & \textbf{13.31} & \textbf{131.74} & \textbf{132.01} & \textbf{109.82} \\
        $\mathcal{P}^2$ & 39.72 & 0.24 & 3.78 & 44.8 & 14.37 & 13.96 & 61.06 & 98.00 & 15.10 & 14.75 & 477.77 & 367.05 & 125.79 \\
        $\mathcal{P}^3$ & 12.88 & 0.22 & 5.61 & 8.34 & 61.65 & 63.62 & 3.30$\cdot 10^7$ & 3.53$\cdot 10^7$ & 51.43 & 50.40 & 2.22$\cdot 10^6$ & 2.19$\cdot 10^6$ & 139.99 \\
        \bottomrule
    \end{tabular}
    \caption{The table shows the results obtained approximating the wind flow in the testing sequences when the parameters were cross-validated and when they were fixed to the optimal.}
    \label{tab:wsvm_fc}
\end{table}